\documentclass[conference]{IEEEtran}
\usepackage{color}
\usepackage{graphicx}
\usepackage{algorithmic}
\usepackage{algorithm}
\usepackage[cmex10]{amsmath}
\usepackage{comment}

\begin{document}
\title{A Memory-Efficient FM-Index Constructor for Next-Generation Sequencing Applications on FPGAs}
\author{\IEEEauthorblockN{Nae-Chyun Chen, Yu-Cheng Li and Yi-Chang Lu}
\IEEEauthorblockA{
    Graduate Institute of Electronics Engineering, National Taiwan University, Taipei, Taiwan, 10617\\
    Email: r04943093@ntu.edu.tw, d01943008@ntu.edu.tw, yiclu@ntu.edu.tw}}
\maketitle
\begin{abstract}
FM-index is an efficient data structure for string search and is widely used in next-generation sequencing (NGS) applications such as sequence alignment and \textit{de novo} assembly.
Recently, FM-indexing is even performed down to the read level, raising a demand of an efficient algorithm for FM-index construction.
%
%Since the construction of FM-index is usually memory-inefficient, it is difficult to implement an FM-index constructor in hardware systems.
%
In this work, we propose a hardware-compatible Self-Aided Incremental Indexing (SAII) algorithm and its hardware architecture.
This novel algorithm builds FM-index with no memory overhead, and the hardware system for realizing the algorithm can be very compact.
Parallel architecture and a special prefetch controller is designed to enhance computational efficiency.
An SAII-based FM-index constructor is implemented on an Altera Stratix V FPGA board.
The presented constructor can support DNA sequences of sizes up to 131,072-bp, which is enough for small-scale references and reads obtained from current major platforms.
%
%Because the proposed constructor needs very few hardware resource, it is suitable to be combined with other FM-index-based applications.
Because the proposed constructor needs very few hardware resource, it can be easily integrated into different hardware accelerators designed for FM-index-based applications.
\end{abstract}


% \fi
\IEEEpeerreviewmaketitle

\section{Introduction}
\label{sec:intro}
Burrows-Wheeler transform (BWT) \cite{burrows1994block} is a string rearrangement algorithm first proposed for data compression.
Making use of the properties between BWT and its original string, FM-index \cite{ferragina2000opportunistic} is designed for efficient string searching.
For a certain string, the data structure of its FM-index contains the BWT, the suffix array (SA) and two auxiliary tables of the target string.
With its high efficiency in both time and memory, FM-index is widely adopted by many next-generation sequencing (NGS) applications \cite{li2009fast, li2010fast, langmead2012fast, simpson2010efficient}.
%
%Sequence aligners such as BWA-backtrack and Bowtie2 utilize FM-index to fast locate reads on references.
%
\par 
For FM-index-based sequence aligners such as BWA-backtrack \cite{li2009fast} and Bowtie2 \cite{langmead2012fast}, the reference sequence is indexed for fast read-locating processes.
For this kind of aligners, because only the reference sequence needs to be indexed, the hardware accelerators  (\cite{olson2012hardware, waidyasooriya2016hardware}) usually build the FM-index externally using CPUs.
%and FM-index constructors on hardware are not needed.
%
Since the length of a reference sequence can be as large as three billion base pairs (bps), the memory usage of naive index constructing method is unaffordable.
%
%Therefore, algorithms in \cite{okanohara2009linear} and \cite{ferragina2012lightweight} are designed to construct FM-index with improved memory usage and speed at the cost of additional working space.
Therefore, how to reduce memory usage in FM-indexing becomes an important issue \cite{okanohara2009linear, ferragina2012lightweight}.
%
%Also, \textit{de novo} assembler such as  also makes use of the index for efficient searching of overlapping regions.
%
\par 
Recently, some \textit{de novo} assemblers (\cite{simpson2010efficient}, for example) apply FM-index for overlap finding of reads.
Also, the reference and reads are both indexed in new sequence aligners such as BWA-SW \cite{li2010fast}.
For these applications, external construction of FM-index has much higher time overhead in comparison with traditional algorithms.
Therefore, on-chip indexing becomes necessary and important.
Our previous research \cite{chen2015power} has demonstrated an FM-index constructor with a lightweight iterative algorithm \cite{ferragina2012lightweight} with an ASIC, but the cost of the indexer is still high when compared to the work proposed here.
%
%Therefore, a memory efficient FM-index construction algorithm suitable for hardware systems is important.
%
\par 
In this work, we propose a novel memory efficient FM-index construction algorithm, Self-Aided Incremental Indexing (SAII), which is suitable for hardware realization.
This algorithm builds the FM-index incrementally.
In each iteration, it utilizes a meta-index to construct the complete index.
Since SAII makes use of the FM-index itself for construction, it has no memory overhead for FM-index-based applications.
Only few computational logic units are needed.
In our hardware system, the processing speed is accelerated by a special prefetch mechanism and a parallel architecture.
We choose Altera Stratix V FPGA as the evaluation platform.
The SAII FM-index constructor is very compact in terms of logic usage, so it can be integrated with other functional blocks to form a complete hardware pipeline in emerging NGS applications.
%and the time performance is acceptable.
%
%This hardware-compatible algorithm provides an effective solution for the hardware acceleration of emerging NGS applications using advanced indexing techniques.
%
%
\section{Background}
\label{sec:background}
\subsection{Burrows-Wheeler Transform}
\begin{figure}[t]
    \centering
    \includegraphics[width=8cm]{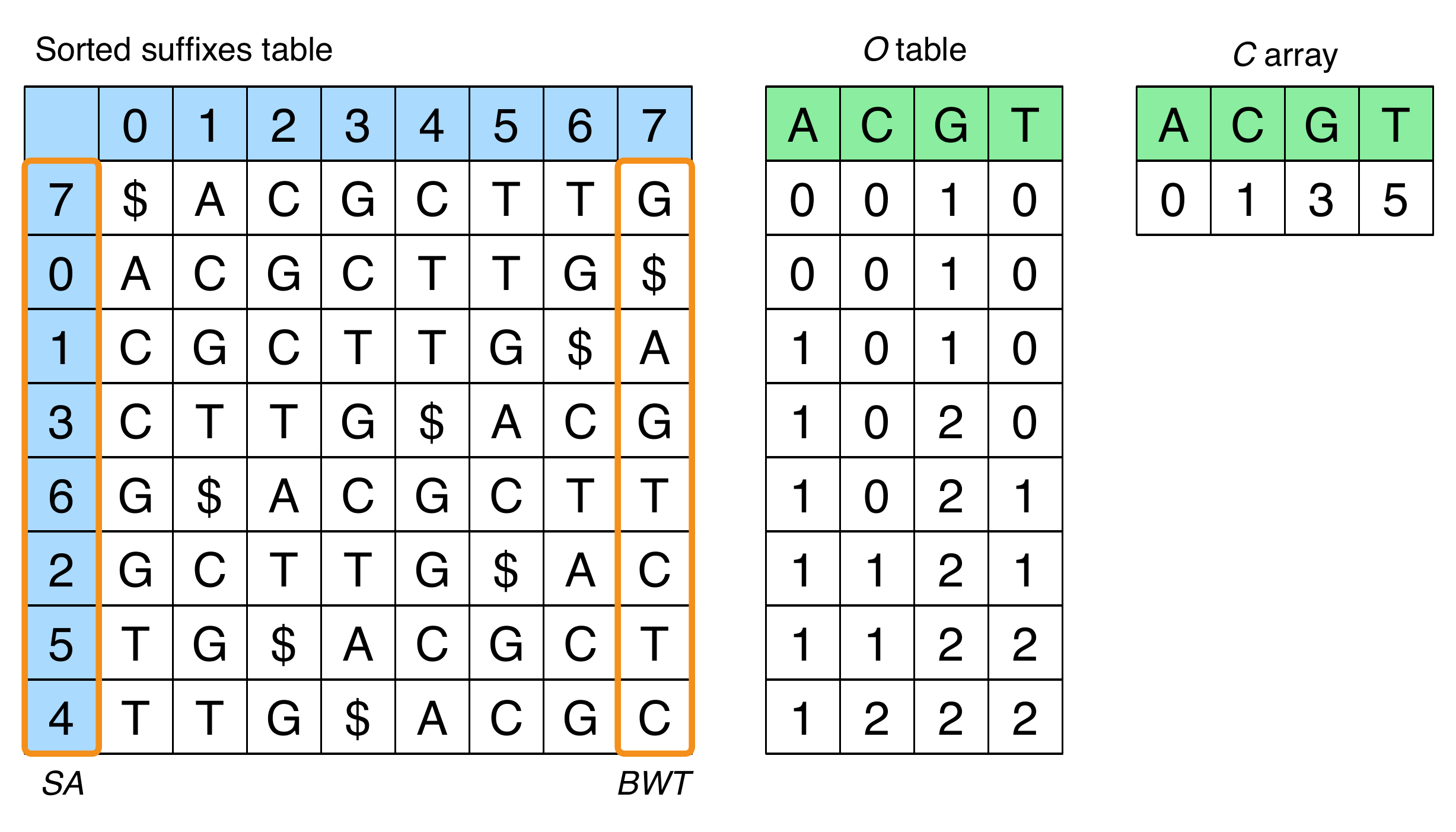}
    \caption{
    	The FM-index of target string ACGCTTG\$.
    	This data structure includes $SA, BWT, C$ array and $O$ table.
    	$BWT$ is the last column of the sorted suffixes table.
    }
    \label{fig:fm_index}
    %\vspace{-10pt}
\end{figure}

%
%
%Burrows-Wheeler transform \cite{burrows1994block} is first used as an effective method to put characters together in a data compression pipeline.
%
To construct the BWT of target sequence $X$, a simple method is via the translation of suffix array ($SA$) with Eq.~\ref{eq:bwt_sa}.
Also, BWT can be obtained by collecting all characters in the last column of sorted suffixes.
Fig.~\ref{fig:fm_index} shows an example of the BWT of sequence $X$=ACGATTG\$, where character \$ is the end-of-string character.
The lexical order is \$$<$A$<$C$<$G$<$T.
%
%With the import of $\sim$, the length of reference sequence can be flexible in hardware systems.
%
%Fig.~\ref{fig:fm_index} also illustrates the naive construction of the FM-index, which requires $O(n^2)$ working space, where $n$ is the length of target sequence.
%
%
\begin{equation}
    BWT[i]=
    \begin{cases}
        X[SA[i]-1]  & \quad \text{if } SA[i]>0 \\
        \$          & \quad \text{if } SA[i]=0
    \end{cases} \label{eq:bwt_sa}
\end{equation}
%
%
% =============================================== %
\hspace{-5mm}
\subsection{FM-index and Backward Search Algorithm}
\label{sec:fm_index}
FM-index is extended from BWT and suffix array, with two auxiliary tables---$C$ array and $O$ table.
The definitions for $C$ array and $O$ table are shown in Eq.~\ref{eq:def_c_array} and \ref{eq:def_o_table}, where $n$ is the length of target sequence $X$.
%
%Count table records the total number of characters lexically smaller than the input character $a$; occurrence table shows the cumulative frequency of the input character $a$ from the starting of BWT to the input index $i$.
%
%Li and Durbin proposed novel heuristics to apply backward search algorithm in sequence alignment, where SNPs and indels increase search time dramatically in traditional approach.
%First, auxiliary array $C(a)$ and table $O(a,i)$ as shown in Eq.~\ref{eq:Ca_def} and \ref{qe:Oai_def} are needed.
%
%
\begin{equation}
C(a)\equiv size\{0\leq j \leq n-2:X[j]<a\}
\label{eq:def_c_array}
\end{equation}
\begin{equation}
O(a,i)\equiv 
\begin{cases}
size\{0\leq j\leq i:BWT[j]=a\}, & i \geq 0.  \\
0, & \text{otherwise.}
\end{cases}
\label{eq:def_o_table}
\end{equation}
\par 
With $C$ array and $O$ table, the position of query $aW$ can be efficiently located within a lower bound $\underline{R}(aW)$ and upper bound $\overline{R}(aW)$ as shown in Eq.~\ref{eq:lower_bound_def} and \ref{eq:upper_bound_def}.
\begin{equation}
\underline{R}(aW)=C(a)+O(a,\underline{R}(W)-1)+1
\label{eq:lower_bound_def}
\end{equation}
\begin{equation}
\overline{R}(aW)=C(a)+O(a,\overline{R}(W))
\label{eq:upper_bound_def}
\end{equation}
\par 
In \cite{ferragina2000opportunistic}, Ferragina and Manzini have proved that $\underline{R}(aW)\leq \overline{R}(aW)$ if and only if $aW$ is a substring of $X$.
This searching algorithm starts from the end of the query sequence and extends iteratively.
Therefore it is also called backward search algorithm.
%
%Originally, FM-index is designed for exact search which allows no mismatches.
%%
%Since the sequences in genomics sequence alignment usually contain mismatches or indels, Li and Durbin \cite{li2009fast} propose novel alogrithms
%%
%%
%With short read length, Backward Search Algorithm can achieve very high processing speed with little loss of accuracy with some heuristics, so it's widely used in novel sequence aligners.

% =============================================== %
% =============================================== %
\section{Self-Aided Incremental Indexing (SAII) Algorithm}
\begin{figure}[h]
    \begin{minipage}{0.49\hsize}    
    \centering
    \includegraphics[width=4cm]{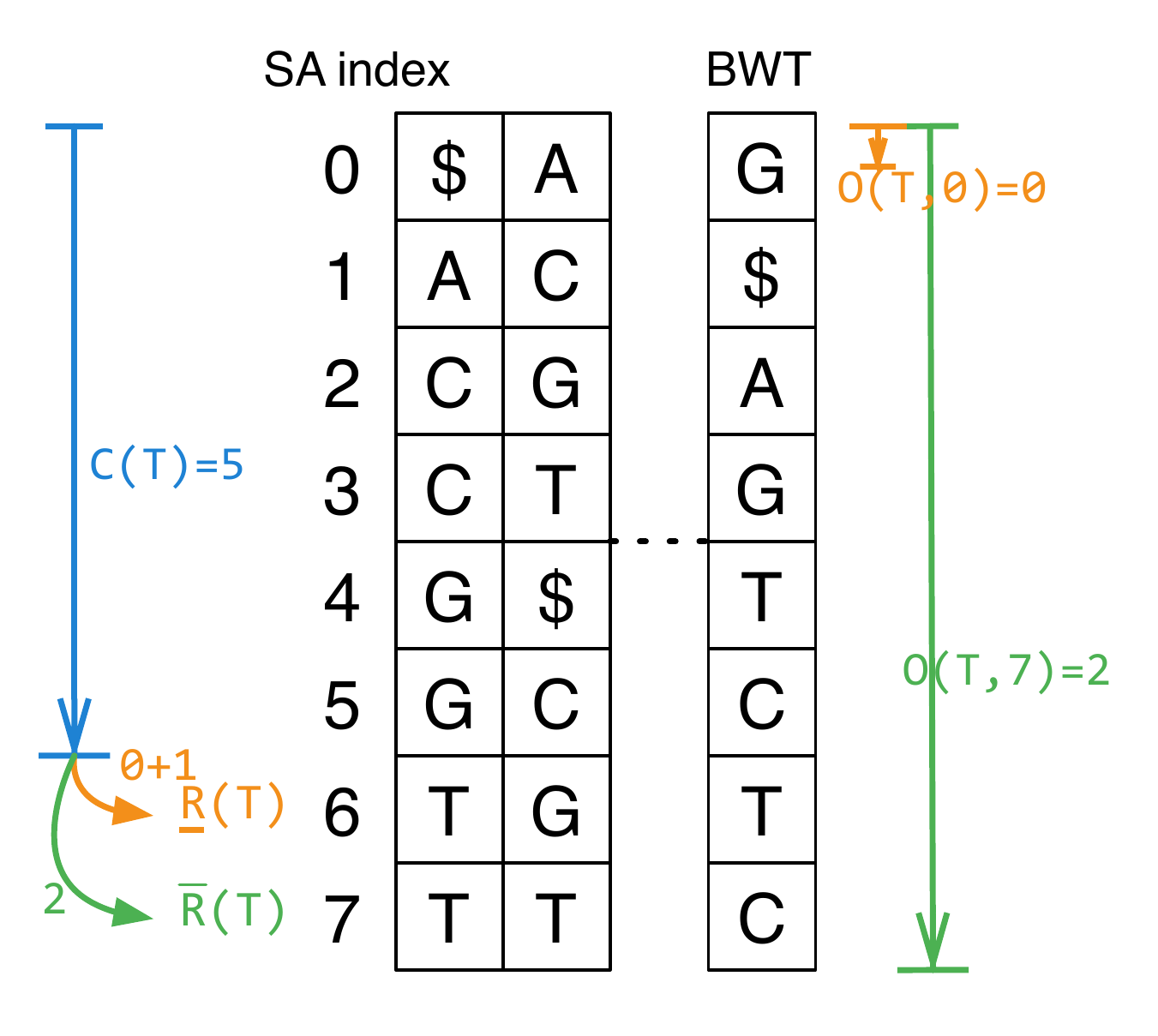}
    \newline \footnotesize{(a)}
    \end{minipage}    
    \begin{minipage}{0.49\hsize}    
    \centering
    \includegraphics[width=4cm]{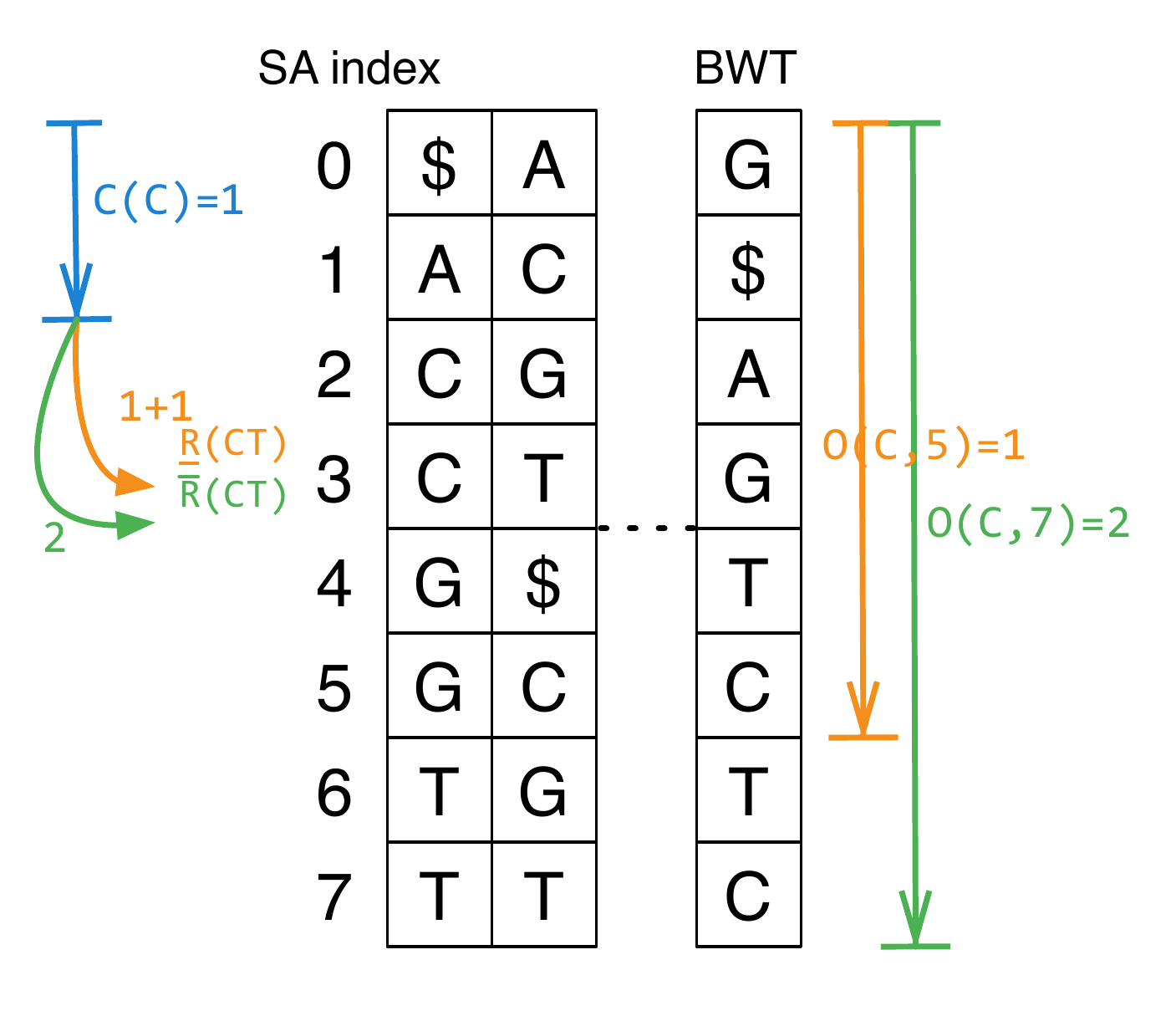}
    \newline \footnotesize{(b)}
    \end{minipage}    
    \caption{An example of searching query sequence CT on indexed target string $X$=ACGCTTG\$. (a) and (b) show the first and second iteration respectively.}
    \label{fig:backward_search}
    %\vspace{-10pt}
\end{figure}

An example of backward search algorithm is shown in Fig.~\ref{fig:backward_search}.
The initial values $\{\underline{R}(\emptyset), \overline{R}(\emptyset)\}$ are set at $\{0, n-1\}$.
Here we discuss the mathematical insights of the lower bound in backward search algorithm.
$\underline{R}(aW)$ is the sum of $C(a)$, $O(a,\underline{R}(W)-1)$ and 1.
$C(a)$ records the total number of characters lexically smaller than $a$ in target sequence $X$.
The $O(a,\underline{R}(W)-1)$ term is the occurrence of $aW$ in $X$.
With an additional offset, $\underline{R}(aW)$ represents the suffix array index of lexically smallest $aW$ sequence.
Similar concept can be used to account for the upper bound.
If $aW$ only occurs once in $X$, $\underline{R}(aW)$ is equal to $\overline{R}(aW)$.
It is also of interest that what would happen if $aW$ is not a substring of $X$.
Since $\underline{R}(aW)$ measures the occurrences of lexically smaller substrings in $X$, the lower bound guarantees the following inequality:
\begin{equation}
suffix_{SA[{\underline{R}(aW)-1}]} < aW < suffix_{SA[{\underline{R}(aW)}]}
\label{eq:not_a_substring}
\end{equation}
\par 
SAII utilizes Eq.~\ref{eq:not_a_substring} to build FM-index incrementally.
For a target sequence $X$, if the FM-index of its substring $L=a_ia_{i+1}...a_{n-2}$\$ has been constructed, the suffix array index of the query string $a_{i-1}L=a_{i-1}a_ia_{i+1}...a_{n-2}$\$ can be determined by calculating $\underline{R}(a_{i-1}L)$.
Since $a_{i-1}L$ could not be a substring of $L$, the suffix array index obtained is unique and follows Eq.~\ref{eq:not_a_substring}.
Therefore, we can insert $a_{i-1}$ into the FM-index of $L$, generating the new FM-index of $a_{i-1}L$ without sorting the whole string all over again.
\par
\begin{figure}[h]
    \centering
    \includegraphics[width=8cm]{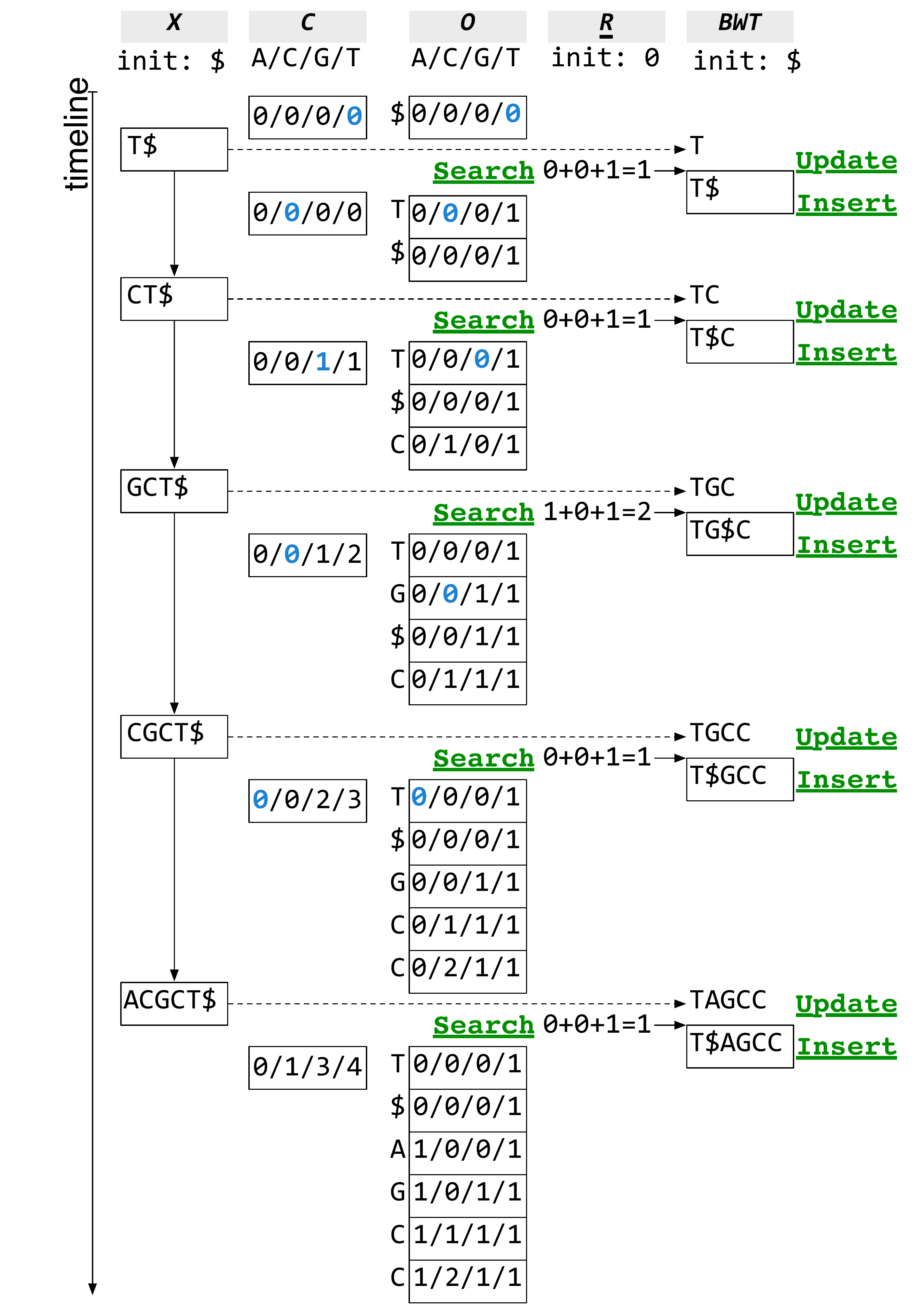}
    \caption{
    An example of the FM-indexing of a target sequence ACGCT\$ with SAII algorithm.
    %
%    In the first iteration, the initial character is $\$$ and the BWT is also $\$$.
    %
%    The corresponding $O$ table and $\underline{R}$ are calculated.
    %
%   In the second iteration, a new character $T$ is added to the target sequence.
    %
%    With the new coming character, the $\$$ in $BWT$ can be immediately updated.
    %
%    With $O$ and $C$ in the previous iteration, the new $\underline{R}$ is calculated and $\$$ is inserted to the $\underline{R}$ position.
    %
% With updated $BWT$, $O$ and $C$ are also updated and then SAII moves on to next iteration.
    %
% After all the characters in the target sequence are read, the corresponding FM-index of the reference is correctly built.
    }
    \label{fig:itr_construct}
\end{figure}

The algorithm of SAII is shown in Alg.~\ref{alg:saii}, and an example of a target sequence ACGCT\$ is provided in Fig.~\ref{fig:itr_construct}.
In the first iteration, the initial character is $\$$ and the BWT is also $\$$.
The corresponding $O$ table and $\underline{R}$ are calculated.
In the second iteration, a new character T is added to the target sequence.
%
%%With the new coming character, the $\$$ in $BWT$ can be immediately updated.
%
Then we use $O$ and $C$ obtained in the previous iteration to calculate the new $\underline{R}$, and the updated $\$$ is inserted to this $\underline{R}$ position to form a new $BWT$.
With this updated $BWT$, we recalculate $O$ and $C$ for this iteration. Then SAII is ready to move on to next iteration.
After all the characters in the target sequence are read, the corresponding FM-index of the reference is correctly built.

Because the construction process is entirely based on FM-index itself, nearly no extra computational resources are needed and the memory overhead is zero for FM-index-based applications.
The time complexity of SAII algorithm is $O(n^2k^{-1})$, where $k$ is completeness of the $O$ table.
The details of $k$ is given in Sec.~\ref{subsec:data_structure}.
\begin{algorithm}
    %\algsetup{linenosize=\small}
    %\scriptsize
    \caption{Self-Aided Incremental Indexing Algorithm}
    \begin{algorithmic}[1]
    \REQUIRE target sequence $X$
    \ENSURE $BWT$, $C$ array and $O$ table
    \STATE Initialize $C$ array and $O$ table
    \STATE $n \gets$ length $(X)$
    \STATE $BWT \gets \$$
    \STATE $q \gets 0$
    \FOR {$i$ from $n-2$ to $0$}
        \STATE $BWT[q] \gets X[i]$
        \label{alg:saii_bwt_update}
        \STATE $L \gets X[i:n-1]$
        \STATE $q \gets C(X[i]) + O(X[i],\underline{R}(L)-1) + 1$
        \STATE $BWT \gets BWT[0:q-1] + \$ + BWT[q:n-1]$
        \label{alg:saii_bwt_insert}
        \STATE Update $C$ array with $L$
        \STATE Update $O$ table with $BWT$
    \ENDFOR
    \end{algorithmic}
    \label{alg:saii}
\end{algorithm}
%
%
% =============================================== %
% =============================================== %
\section{Hardware Implementation and Discussion}
\label{sec:hardware}
\subsection{Overall Architecture}
\begin{figure}[t]
	\centering
	\includegraphics[width=7cm, clip]{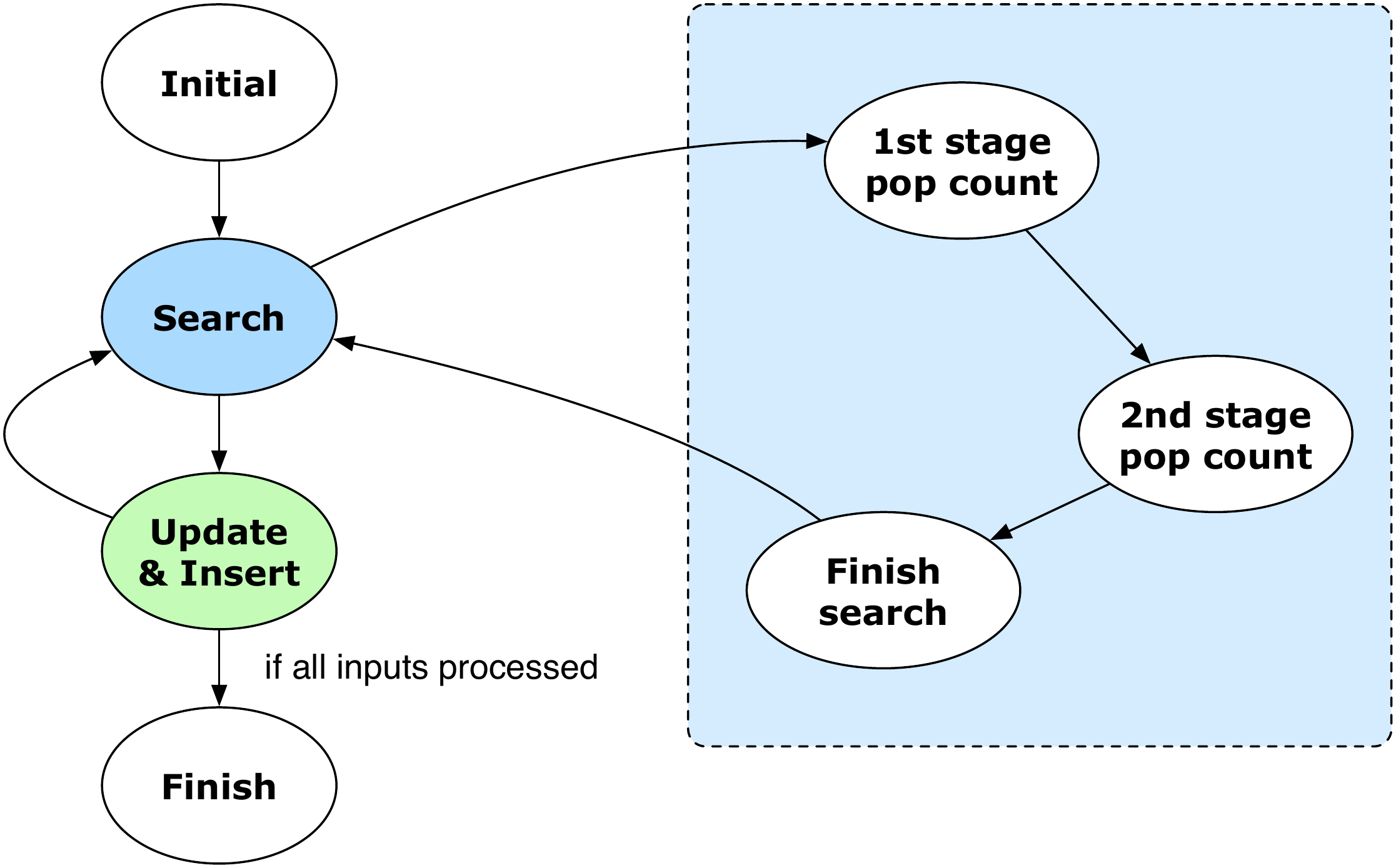}
	\caption{The finite-state machine controlling the hardware system.}
	\label{fig:FSM}
    %\vspace{-10pt}
\end{figure}

The hardware system includes a finite-state machine controller and a combinational computing logic.
The finite-state machine of the proposed hardware is shown in Fig.~\ref{fig:FSM}.
There are four states in this system: Initial, Search, Update and Finish states.
\par 
Initial and Finish states control the initial and finishing conditions of the hardware.
Search state computes the lower bound $\underline{R}(aW)$ with Eq.~\ref{eq:lower_bound_def}.
A two-stage parallel pop counter is used in Search state for fast computing.
Update State updates the latest incoming symbol to the \$ position in previous iteration.
Also, it inserts the \$ symbol to the updated index based on the position calculated in Search state.
The whole FM-index data structure including $BWT$, $C$ array and $O$ table all need to be updated in this state.
The finite-state machine repeats between Search and Update states to construct the FM-index and moves to Finish state after the last character of the target sequence is processed.
%
%
% =============================================== %
\subsection{Data Structure}
\label{subsec:data_structure}
For DNA aligners, the size of alphabets ($\Sigma$) is five (including \$).
Since symbol \$ occurs only once, in our hardware system only A,C,G, and T are encoded.
End-of-string character \$ is encoded the same as the symbol, A, and an additional special pointer is designed to store the position of \$.
This design uses only two bits for each character, which is only 67\% in comparison to that of naive 3-bit encoding.
\par 
%For FM-indexed-based hardware systems, usually memory uses lots of resources \cite{chen2015power}, so it's very important to allocate the memory properly.
%
The memory usage of the three main components of FM-index, $C$ array, $O$ table and $BWT$, are $4\log{n}$, $4n\log{n}$ and $2n$, respectively.
However, the length of genome sequence data is sometimes very large.
The length of a human chromosome can exceed 200 Mbp and the whole human genome is more than 3 Gbp.
With this scale of data, the $4n\log{n}$ memory usage of a complete $O$ table is very expensive in hardware systems.
It should be noted that $O$ table is a hash table obtained from $BWT$ designed for fast computation of $\underline{R}$ and $\overline{R}$.
The correct bounds can still be calculated even without $O$ table at the cost of searching efficiency.
\par 
In our hardware system, incomplete $O$ table is used to achieve balance between memory usage and computing speed.
An incomplete $O$ table stores the occurrence values at every $k$ characters.
It is $k$ times smaller than a complete table.
With an incomplete $O$ table, the calculation of $O(a,i)$ is split into two parts.
First, $O(a,k\left\lfloor{\frac{i}{k}}\right\rfloor)$ is stored in the incomplete table.
Second, the occurrences of $a$ from $k\left\lfloor{\frac{i}{k}}\right\rfloor$ to $i$ is calculated with a pop counter.
%
% and the computing time for counting is at most $k$.
%%
%Also, searching is accelerated with a parallel pop counter design.
%
In our hardware implementation, $k$ is set at 2,048.
The pop counter is designed with a two-stage architecture, in which the first stage has 32 parallel adders and the second stage has 64 parallel adders.
The incomplete $O$ table can be adjusted for different applications with simple modification of parameters.
%
% =============================================== %
\subsection{Constructing BWT with Prefetch Mechanism}
\begin{figure*}[t]
	\begin{minipage}{0.59\hsize}
	\begin{minipage}{1\hsize}
	\centering
	\includegraphics[width=9cm, clip]{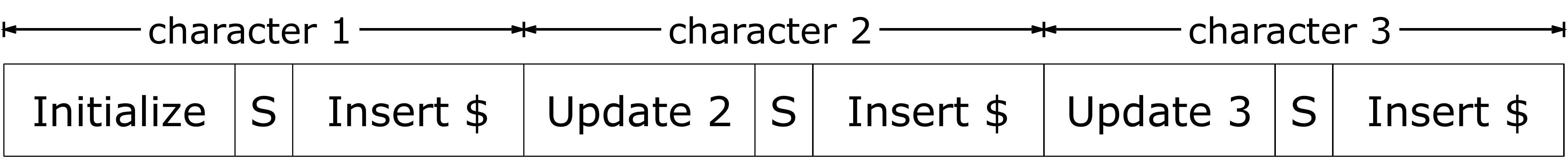} 
    \newline     
     \footnotesize{(a)}
	\vspace*{3mm}
    \end{minipage}
    \newline
    \begin{minipage}	{1\hsize}
	\centering
	\hspace{-2mm}
	\includegraphics[width=9cm, clip]{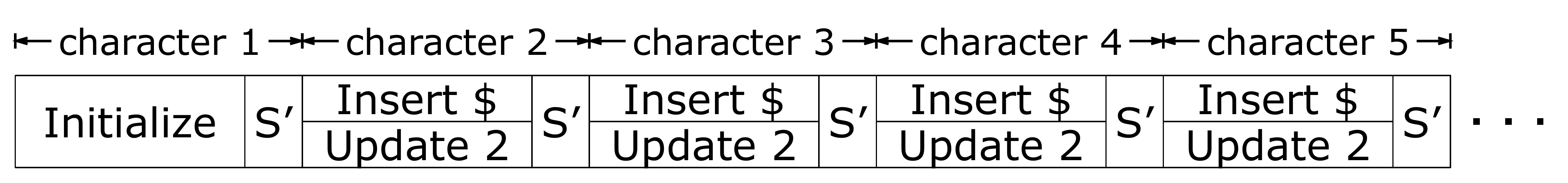}
    \newline
    \footnotesize{(b)}
    \end{minipage}	
	\vspace*{0.5mm}
	\caption{
		%Assume that one update takes $q$ cycles and one backward search takes $m$ cycles ($q>>m$), prefetch mechanism reduces the computing time from $2q+m$ to $q+m$ for each iteration.
         Timing diagrams for: (a) constructing BWT without prefetch mechanism, and (b) constructing BWT with prefetch mechanism.
		The blue boxes (S- and S'-boxes) denote Search state.
		The S'-boxes contains the additional monitor.
	}
	\label{fig:early_update}
	\end{minipage}
    \hspace{2mm}	
	\begin{minipage}{0.39\hsize}
    	%\begin{figure}
	\centering
	\includegraphics[width=6cm]{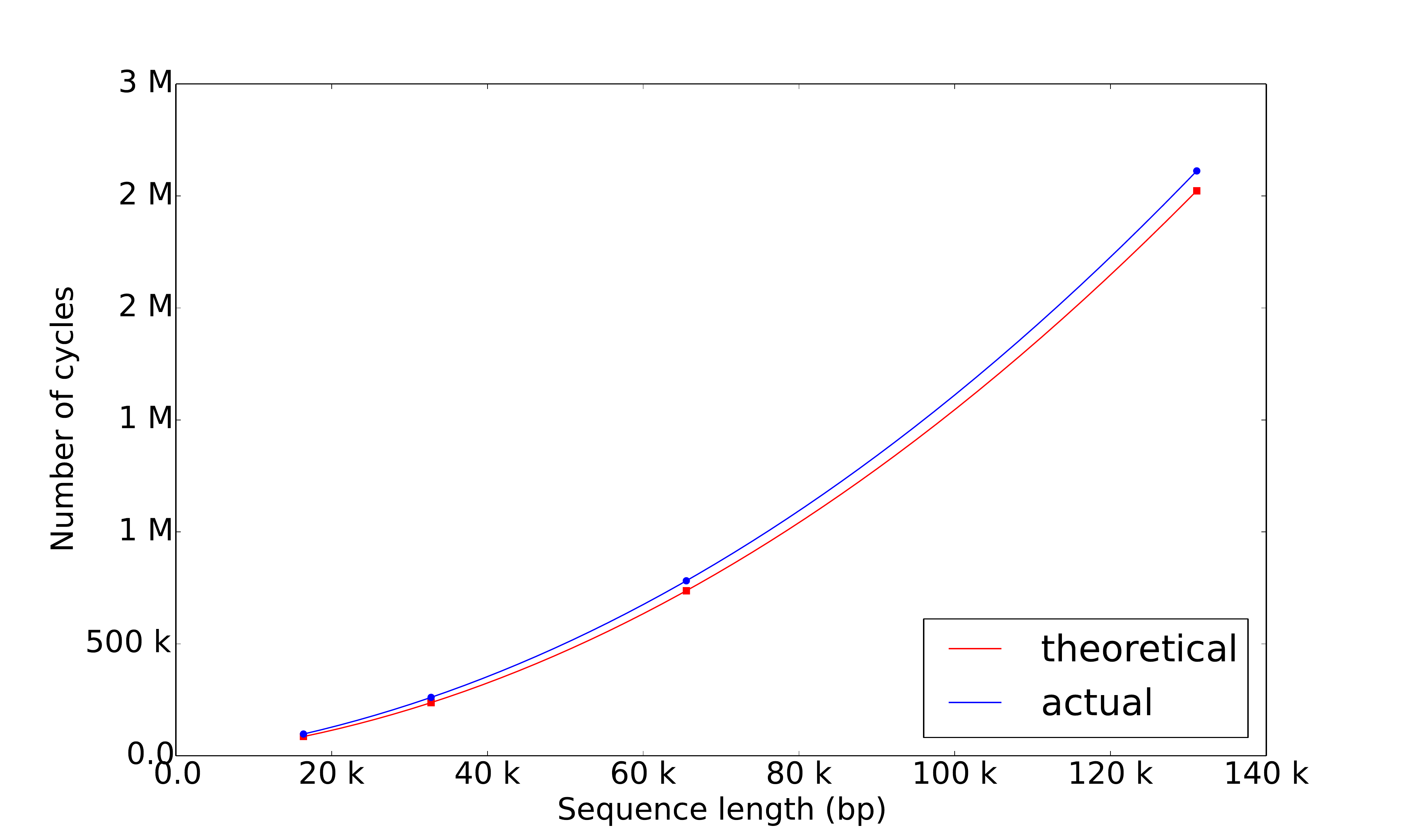}
	\caption{Processing cycle count of SAII hardware system.}
	\label{fig:runtime}
    %\vspace{-10pt}
    %\end{figure}
	\end{minipage}
\end{figure*}

In our hardware system, Search and Update states are in charge of the construction of FM-index.
To save computing resource, $BWT$ and $O$ table are both segmented and stored in the BRAM.
Though this saves lots of area, it takes longer time to update the BRAM due to the fixed word length.
Therefore, how to make use of the limited bandwidth is very important.
To address this issue, we design a prefetch mechanism that saves 50\% time of BRAM updating.
The timing diagram without prefetching is shown in Fig.~\ref{fig:early_update}(a).
%
%In this example, we assume full size $O$ table is used; if the $O$ table is compressed, the early update mechanism works similarly.
%Once the system enters construction stage, it circulates between Search-Update and Insert states.
%
In Search state, $\underline{R}$ is calculated; in Update state, the incoming character is updated to the FM-index as shown in Line~\ref{alg:saii_bwt_update} in Alg.~\ref{alg:saii}; in Insert state, the $\$$ is inserted to the FM-index, as shown in Line~\ref{alg:saii_bwt_insert} in Alg.~\ref{alg:saii}.
In both Update and Insert states, $BWT$ and $O$ table in the BRAM have to be refreshed, so the processing time is long.
%
%Also, the $\$$ in BWT is updated with new symbol.
%In Insertion State, new $\$$ is inserted to the correct position, which is $\underline{R}$, of BWT.
%The updating of new symbol in Search-Update State and the insertion of new $\$$ in Insertion State all cost $O(n)$ time.
%
%\par 
\par
Prefetch mechanism is designed to reduce the runtime to refresh BRAM.
As shown in Fig.~\ref{fig:itr_construct}, SAII algorithm first replaces the $\$$ in $BWT$ with the new character, calculates the new insertion position of $\$$ and inserts new $\$$ to the $BWT$.
With prefetch mechanism, the insertion of $\$$ is combined with next Update state and executed after the hardware system sees the next character.
This does not generate the completely correct FM-index yet because of its early update design.
Therefore, the early updated position and character have to tracked with an additional monitor to make sure the calculation of $\underline{R}$ is correct in the next iteration.
In the last iteration, since there is no more character for prefetching, the final FM-index is correct.
Fig.~\ref{fig:early_update}(b) shows the timing diagram of the hardware system with prefetching.
\par
Assume that one update takes $q$ cycles and one backward search takes $m$ cycles ($q>>m$), prefetch mechanism reduces the computing time from $2q+m$ to $q+m$ for each iteration.
The computing time is nearly half of the original design.
The expected runtime ($\mathcal T$) of the SAII hardware system is shown in Eq.~\ref{eq:timing_model}, where $m$ stands for search time and is set to 3 cycles in our implementation.
\begin{equation}
\mathcal T = k\sum^{\frac{n}{k}}_{i=1}(m+\cfrac{i}{2})
\label{eq:timing_model}
\end{equation}

%With early update, Insertion State doesn't update the BWT with new $\$$ symbol.
%Instead, it requests the next symbol and inserts it in BWT.
%This changes the value of $O$ table but it can be fixed with simple logic during the computing of $\underline{R}$.
%Therefore, the total updating time is greatly reduced.
%

%
%\par 
%
%
% =============================================== %
% =============================================== %
\subsection{Discussion}
\label{subsec:discussion}
We implement our SAII FM-index constructor on an Altera Stratix V FPGA (5SGXEA7N2F45C2N) board.
The hardware system is synthesized using Altera Quartus (v.15.0) tool.
It only uses 21,944 ALMs (9\%) and 266,496 BRAMs ($<$ 1\%) on this FPGA.
Also, the serial-input design uses a very small proportion of the I/O bandwidth.
These properties make the SAII algorithm easily integrated into existing sequencing pipelines at very low hardware cost.
\par 
The operation frequency can reach 120 MHz even with the worst case model (900 mV, 85 $^{\circ}$C).
Sequences with different lengths, from 16,384-bp to 131,072-bp, are tested and the results are shown in Fig.~\ref{fig:runtime}.
It takes about 21 ms to finish the indexing of a 131,072-bp sequence.
The runtime is very close to our theoretical estimation given in Eq.~\ref{eq:timing_model}.
Even for genomes with several million base pairs, it is estimated that our system can construct the FM-index in seconds.
%
%
%\begin{table}[t]
%\centering
%\caption{Hardware Resources Used on Altera Stratix V FPGA}
%\label{table:fpga_synthesis}
%\begin{tabular}{ll}
%Logic utilization (in ALMs) & 21,944 (9\%) \\ \hline
%Total registers & 4780 \\ \hline
%Total block memory units & 266,496 (\textless1\%) \\
%\end{tabular}
%\end{table}
%
% =============================================== %
% =============================================== %
\vspace{2.5mm}
\section{Conclusions}
\label{sec:conclusions}
With many emerging applications based on FM-index, an efficient index construction algorithm is needed.
Previous algorithms (\cite{okanohara2009linear, ferragina2012lightweight}) need additional working space to build the index, raising the costs of hardware systems.
In this paper, we propose a novel hardware-compatible Self-Aided Incremental Indexing (SAII) algorithm to construct FM-index with no memory overhead.
This algorithm is accelerated with a parallel pop counter and a special prefetch mechanism.
Its realization on FPGA needs very few hardware resources and can be easily integrated in different FM-index-based applications.
%
%\vspace{5mm}
%
%
\section*{Acknowledgment}
This work is supported by the Ministry of Science and Technology, Taiwan, under Grant numbers MOST 105-2221-E-002-090 and 106-2221-E-002-055. Nae-Chyun Chen would like to thank NOVATEK for providing fellowship.
%
%\vspace{5mm}
%
\bibliographystyle{IEEEtran}
\bibliography{IEEEabrv,paper}
\end{document}